\documentclass[12pt]{spieman}  % 12pt font required by SPIE;
\usepackage{amsmath,amsfonts,amssymb}
\usepackage{graphicx}
\usepackage{setspace}
\usepackage{csquotes}
\usepackage{tocloft}
\usepackage{hyperref}
\usepackage{pdfpages}

\title{Defining the quantum workforce landscape: a review of global quantum education initiatives}

\author{Maninder Kaur}
\author{Araceli Venegas-Gomez}
\affil{QURECA (Quantum Resources and Careers), 272 Bath Street, Glasgow, Scotland, G2 4JR, United Kingdom}

\cftpagenumbersoff{figure}

\begin{document} 
\maketitle

\begin{abstract}
Rapid advances in quantum technology have exacerbated the shortage of a diverse, inclusive, and sustainable quantum workforce. National governments and industries are developing strategies for education, training, and workforce development to accelerate the commercialization of quantum technologies. In this paper, we report the existing state of the quantum workforce as well as several learning pathways to nurture the talent pipeline between academia and industry. We provide a comprehensive guide of various educational initiatives accessible throughout the world, such as online courses, conferences, seminars, games, and community-focused networks, that facilitate quantum training and upskill the talent needed to develop a better quantum future.    
\end{abstract}

% Include a list of up to six keywords after the abstract
\keywords{quantum education, quantum workforce, quantum technologies}

% Include email contact information for corresponding author
{\noindent \footnotesize\textbf{*}Maninder Kaur,  \linkable{maninder.kaur@qureca.com} }

\begin{spacing}{2}   % use double spacing for rest of manuscript

\section{Introduction}\label{section1}  % \label{} allows reference to this section
Quantum technologies, the next cutting-edge technological revolution, have seen a significant shift from mostly scientific research, hardware approaches and experimentation to practical product and commercialization in recent years \cite{McKinsey, QComm}. By exploiting the quantum-mechanical phenomena such as superposition, interference, or entanglement to perform computations, quantum computers will be able to efficiently solve hard computational problems that today’s classical computers cannot solve. Although the state-of-the-art fault-tolerant quantum computer is still in its infancy\cite{preskill}, the potential impacts of quantum technologies have been recognized worldwide and many countries are making efforts in initiating national quantum strategies to foster its development within the global landscape \cite{Quaninit}. 

Rapid advancements in quantum technologies are enabling organizations across a range of industries adapting practices in their daily operations to attain a quantum advantage and remain agile in a shifting global economy \cite{PhysRevResearch.4.013006, QTraffic, QInternet}. However, these organizations are facing a major obstacle in their journey to quantum advantage, being recognized as a “quantum bottleneck”, the shortage of well-trained academics and professional talent specialized in quantum science and technologies \cite{Guardian}. The demand for the quantum-ready workforce has increased considerably with the development of various roles in a diverse array of sectors and applications \cite{Techshort18}.  Therefore, the need to address the challenge of the quantum workforce education is gaining an incredible traction across the world. 

National governments and commercial organizations worldwide are formulating critical strategies that will eventually underlie the burgeoning quantum economy. The United States Government outlined four key strategies in its Quantum Information and Technology Workforce Development Plan to address the expansion of the quantum workforce \cite{qistUS}. One of the main strategies has been focused on broadening student participation and accessibility to post-secondary education, including associate's, bachelor's, master's and professional training programs with a focus on quantum skills at all levels to address the workforce gaps. To boost the European Quantum Industry, the European Quantum Industry Consortium (QuIC) has created a strategic industrial road map to shape the education and training of the European quantum-ready workforce \cite{QuIC}. Regional industries consortium such as the Quantum Strategy Institute (QSI) \cite{QSI}, Quantum Industry Canada (QIC) \cite{QIC}, the Quantum Technology and Application Consortium (QUTAC) \cite{QUTAC}, and the Quantum Ecosystems Technology Council of India (QETCI) \cite{QETCI} are identifying strengths and gaps in the current quantum ecosystem's existing education and workforce development. The Quantum Flagship’s Quantum Technology Education Coordination and Support Action (QTEdu CSA) intends to bridge the gap between the academic and industrial quantum communities and is currently running 11 pilot projects in 25 European countries to address educational needs and inform society about quantum technologies \cite{QTEdu}. The UK National Quantum Technology Programme has included training and skills as a key objective in its strategic plan to raise the profile of quantum technologies in educational curricula for building the future workforce pipeline \cite{UKQU}. The Sydney Quantum Academy (SQA) \cite{SQA}, Canada's National Quantum Strategy \cite{NQSC}, the Singapore’s Quantum Engineering Program \cite{QEPS} and India's National Mission on Quantum Technologies and Applications \cite{QTInd} are among other initiatives who address the issue of education and training at all levels to establish a diverse talent with quantum expertise. 

Designing and implementing a quantum workforce plan is extremely challenging because this area has not been fully investigated and is still unknown \cite{Hilton19}. With the growing number of start-ups, companies, universities, large corporations, and government agencies that are accelerating the quantum technologies industry, the problem of skills shortage on both, demand and supply side, cannot be ignored. A continuous expansion of the quantum ecosystem demands the growth of the quantum talent pool who can understand the core concepts in Quantum Information Science and Technology (QIST), and address challenges such as scalability and identification of unique real-world opportunities for quantum systems \cite{Araceli20}. In order to meet the anticipated industrial demand for a quantum-ready workforce, several programs have been established (or will be in the near term), including undergraduate \cite{asfaw2021building}, master’s and PhD level academic degrees \cite{PhysRevPhysEducRes.16.020131}, industrial educational programs \cite{Qiskit21,Azure21,Google21}, online courses \cite{EdX22, Coursera, Qureca22}, and community focused networks \cite{qworld22,onequantum22}.

The primary objective of this paper is to present an overview about the education and training resources available worldwide which are key to develop the current and the future quantum workforce. Furthermore, we describe potential resources and activities to train that workforce.
The rest of this article is organized as follows: in Sec.~{\ref{section2}}, we describe the methodology employed in our research, as well as certain limitations in gathering and analysing the data presented in the paper. In Sec.~{\ref{section3}}, we analyze the current quantum technologies job market emphasizing the importance of general degree qualifications and professional experience for numerous job opportunities. This research will be beneficial in developing a strategy for a career in the quantum industry, as there are several ways to acquiring quantum skills and training. The potential paths for academic graduates leading to employment in the quantum industry are discussed in Sec.~{\ref{section4}}. In Sec.~{\ref{section5}}, we describe the current learning and training resources in the field of quantum technologies, such as academic degree programs, online education, community events and seminars, games, and other general tools to support the workforce development. In Sec.~{\ref{section6}} we present a general strategy for industry to develop a skilled workforce, before providing a summary and conclusion highlighting the significance of retraining and upskilling the current talent, as well as the need to promote the awareness of global quantum education, in Sec.~{\ref{section7}}.

\section{Methodology}\label{section2}
To overcome the challenges of building a viable quantum-ready workforce\cite{qistUS, Qsmart}, we address three main questions in this study - (1) identify and assess the current quantum ecosystem's competences and education needs, (2) determine the optimal path to bridge the gap between academia and industry, and most importantly, (3) where to find all the learning and training resources to start a career in quantum technologies.

To analyze the global quantum technologies market demand, the Quantum Computing Report jobs database was accessed for collecting data on commercial job vacancies \cite{QCR22}. The database search excluded job vacancies from academic and non-profit organizations, and was limited to job vacancies in the quantum companies. Depending on their main field of operations, these companies were further categorised in sectors such as quantum computing (hardware and software), quantum communications, quantum security, quantum sensing \& imaging, enabling technologies, and consulting \cite{QTDomains}. To eliminate biases in the analysis, we evaluated the minimal qualifying criteria of degrees as well as professional experiences. It can be argued that the future quantum workforce will require a broad range of skills over specialized STEM degrees \cite{hughes2021assessing}. Although it would be insightful to observe the importance of skills demand for various job vacancies, this assessment is beyond the scope of this paper.

Additionally, to address the question of finding resources for quantum education and training, we reviewed all quantum initiatives currently active around the world. We gradually gathered all the education initiatives through publicly available resources, web searches, journal publications, community events or conferences, and academic and industrial collaboration during the period of 2019-2022. This paper serves as a guide for all those learners, students, researchers, or industry experts who are seeking educational tools and resources to help them prepare for a career in the field of quantum technologies.

As the main limitation of our study, although we believe it is highly representative, we cannot argue that it is exhaustive. Thus, it should be noted that our overview provides a summary at the time this paper is written, and additional resources will appear in the future. 

In the next sections, we present the key outcomes of our analysis.

\section{Quantum technologies market demand}\label{section3}

\begin{figure}
\begin{center}

\includegraphics[height=9cm]{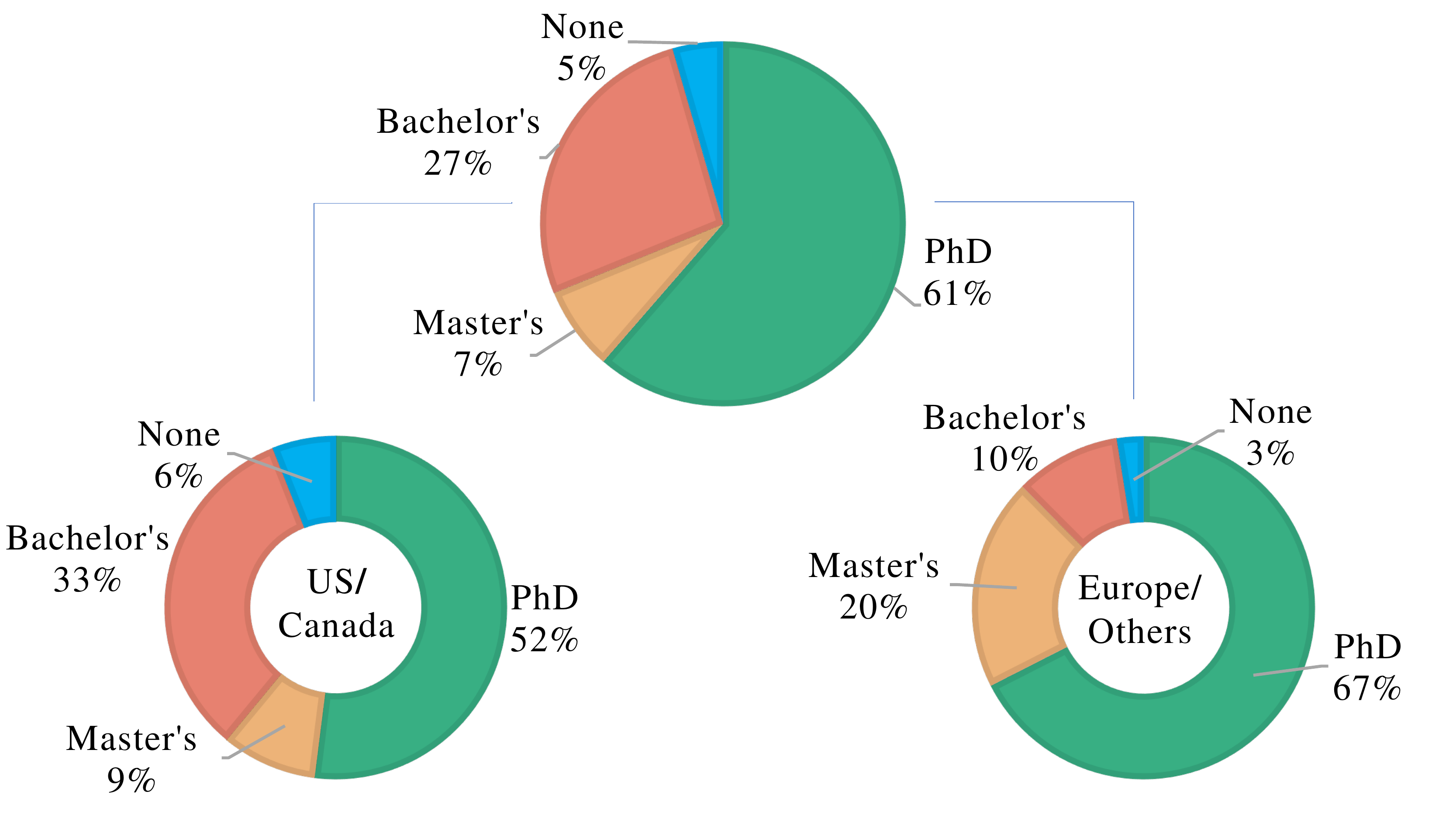}

\end{center}
\caption 
{ \label{fig:1}
Degree requirements for various job vacancies across (a) the world, and in (b) the US/Canada and (c) Europe/rest of the world.\cite{QCR22}}
\end{figure}

As the field of quantum technologies is accelerating, the demand for highly advanced knowledge and skills in quantum will only grow by time \cite{Quaninit, Araceli20, Qsmart}. There are distinct workforce requirements for its development, implementation in real-world applications, and commercialization. To assess the quantum workforce landscape, we examined the current estimated job market and their academic and work experience requirements for quantum professions \cite{QCR22}. Fig.~\ref{fig:1} shows the distribution of academic degree (PhD, master's, bachelor's, or no degree) requirements for the job vacancies across the world, which are further segmented by two regions: US/Canada, and Europe and other countries. It can be seen that graduates with PhD degrees are by far the most in demand, followed by bachelor's and master's graduates. We investigated over 750 vacancies currently available of which 62 percent demand the candidates for a PhD degree. Furthermore, the requirement of higher education for job vacancies is significantly stronger in Europe than in the US and Canada. Surprisingly, the majority of the job vacancies demand for bachelor's graduates in US/Canada compared to Europe/rest. 

\begin{figure}
\begin{center}

\includegraphics[height= 6cm]{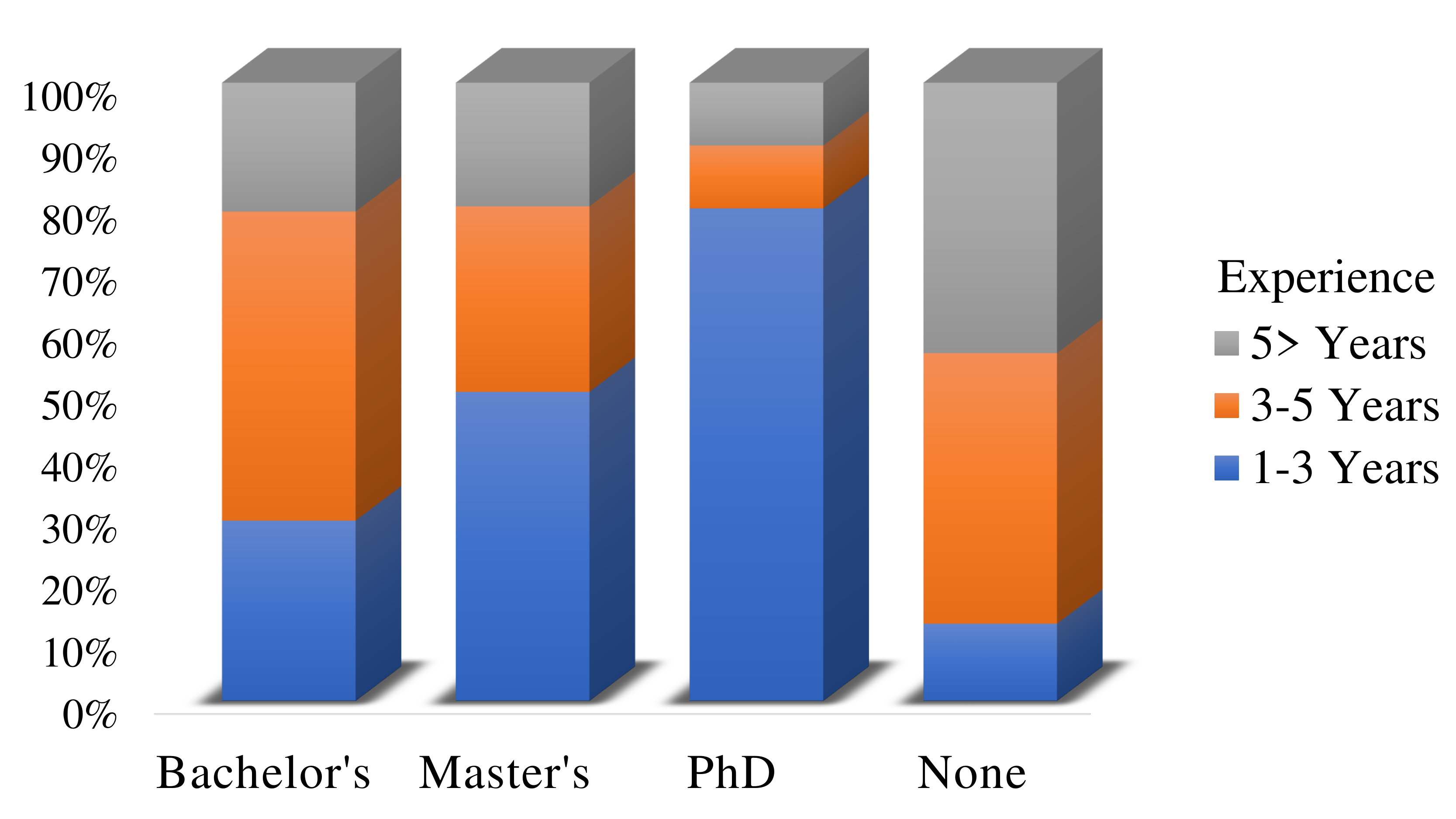}

\end{center}
\caption 
{ \label{fig:2}
Requirements of minimum professional experience in addition to degrees for various job vacancies available across the world. \cite{QCR22} }
\end{figure}

Furthermore, we analyzed the number of years of work experience demanded to fill in these vacancies. Fig.~\ref{fig:2} shows the advertised positions with their required academic degrees and minimum professional experience in years. There is clearly a large divergence between PhD degrees and bachelor's degrees. While bachelor's graduates having professional experience may be crucial for the roles which requires the basic quantum knowledge attained through one or two courses, the criteria for professional experience needed for job vacancies are comparatively lower for PhD graduates who have been trained in specialized quantum capabilities. It should be noted that this analysis do not focus on skills needs, but rather on academic degree requirements. It is often argued whether the academic degrees train the graduates in industry specific skills and satisfy to the needs of workforce in future. Following discussions with stakeholders in the quantum industry, there is a consensus that when graduates acquire industry training through internships or other technical, business, and soft skills through community engagements, then they become more valuable candidates in the employment market.

It is noteworthy to mention that approximately 37\% of quantum-based roles in the current job market are from large corporations (such as IBM, Google, Microsoft, Quantinuum, Amazon, and IonQ), the majority of which have their headquarters in the US/Canada. These companies are hiring expertise either with a bachelor's degree having 3-5 years of work experience or advanced degrees (masters or PhD) with 0-1 years of experience. This factor contributes towards the higher demand of bachelor's graduates in US/Canada. In addition, it is found that the current hiring needs are focused in the global quantum computing sector (42\%), followed by companies specialized in quantum hardware (32\%) and quantum software (13\%). There has been an increase in hiring across other sectors such as enabling technologies (5\%), quantum security (4\%), quantum communications (3\%) and quantum sensing \& imaging (2\%). Moreover, to develop a flexible, well-trained, and multidisciplinary workforce, these companies also require graduates with expertise in a wide range of disciplines including physics, materials science, electrical engineering, computer science, chemistry, and mathematics.

\section{Broadening career pathways}\label{section4}

The most important aspects of any educational program are the development of fundamental concepts and expertise required for future learning and the ability to demonstrate the skills sought by prospective industrial employers for various roles \cite{PhysRevPhysEducRes.16.020131}. It is imperative to say that the current linear learning path pursued by graduates to find a career in quantum-related industries, i.e., earning a PhD with a specialization in QIST and/or research experience gained in a related field, will eventually not be the only path to satisfy the requirements of a quantum-ready workforce. The academic educational programs should be designed with multiple entry points with the prospect of building a satisfying career in quantum computing, for example - pre-college exposure, undergraduate, graduate degrees and field training, and certificate programs for postgraduates \cite{OSA2}.

\begin{figure}
\begin{center}

\includegraphics[height=9cm]{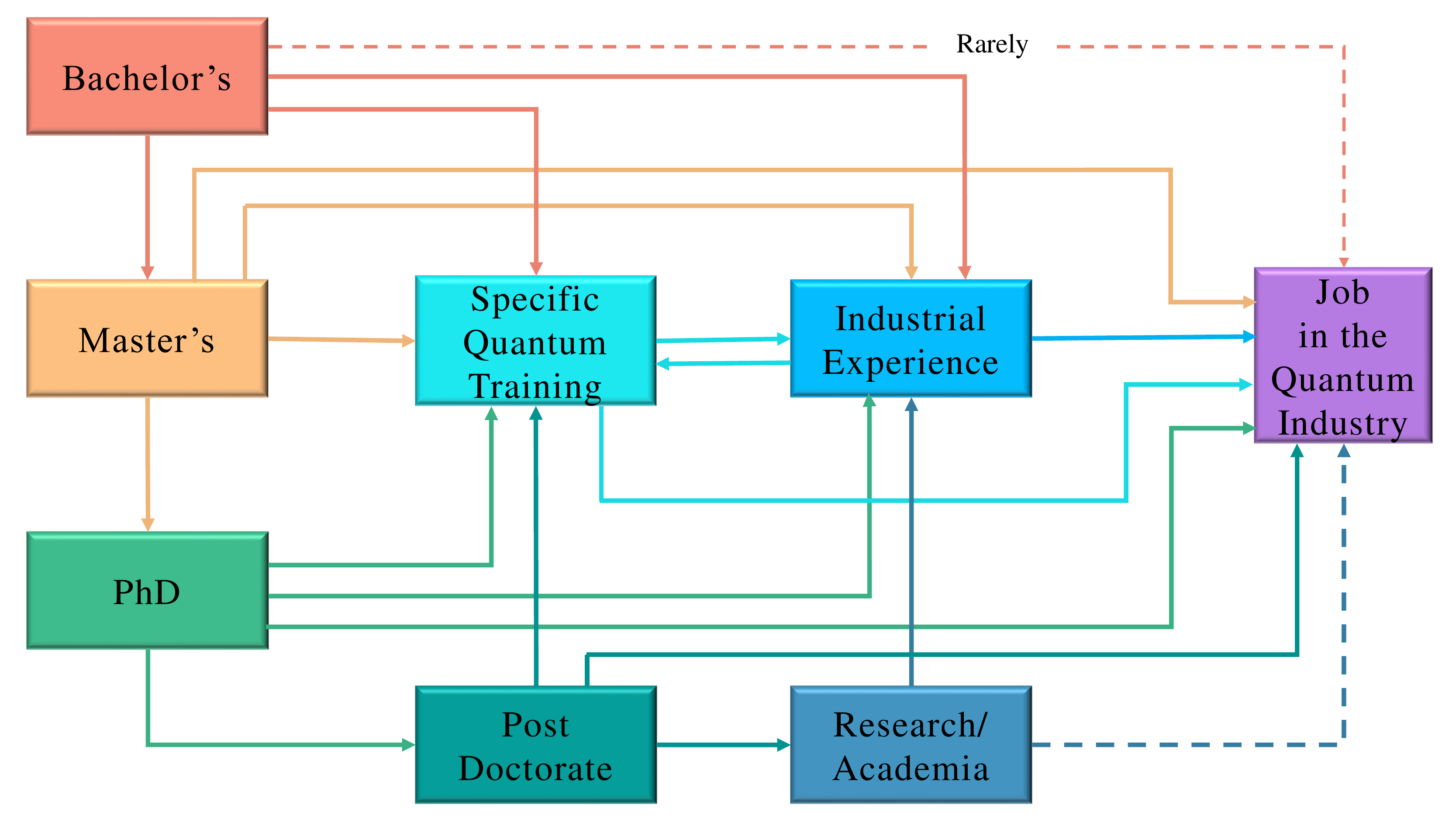}

\end{center}
\caption 
{\label{fig:3}
Career pathways for academic graduates leading to employment in the quantum industry.} 
\end{figure}

When it comes to choosing the best route to start a career in quantum technologies, it is highly unlikely to get a job right after completing a bachelor's degree program, as shown in Fig.~\ref{fig:2}. While employers place a high value on essential technical abilities (such as knowledge in science, engineering, math, or coding) and soft skills (such as communication, decision-making, and problem solving), they are more likely to invest in candidates having industry-specific experience or job-specific skills. This can be achieved either through advanced master's programs or specific industrial training in quantum technologies. Fig.~\ref{fig:3} shows a broad spectrum of different pathways that facilitate quantum education and training to students at all levels and lead them to find a job in the quantum industry. Combining traditional learning with specialized academic disciplines and shorter-term hands-on training with real-world projects will be critical in developing the skills needed for in-demand jobs. It is much easier to gain a pathway to a quantum industry job if you have had specific quantum training achieved through various online courses, workshops, hackathons, and summer schools, and/or industry experience attained through mentorship or internship programs. This adaptable approach to train learners at several entrance points enables them to transfer rapidly into the labour market. That is why the plethora of current educational resources range from specialised academic programs to specific courses for industry professionals (described more in detail in the next section).

Fig.~\ref{fig:3} explicitly shows these different pathways which can help to improve the gap of transferable skills mentioned. It is possible to gain access to a job in quantum industry directly from education if it is pursued within the quantum technology field of research directly linked to the job requirement. Understanding these alternative career pathways is useful, because with specific quantum training, many more potential employees will be equipped with the adequate quantum skills for the industry, filling the requirements from the job market, as analyzed in Sec.~{\ref{section3}}. The right quantum education initiatives can indeed play an important role in bridging the gap between the academia and the industry. The overview of these initiatives is presented in the next section.

\section{Quantum education initiatives}\label{section5}

Building a trained quantum workforce needs a holistic approach to skill development that includes broad access to quality education, a pipeline between education providers and industry, continuous workplace training, and the anticipation of future industrial competencies needs. Numerous efforts are currently underway to provide a wide range of education and training options to support career transitions and workforce deployment, and to increase the resilience of those currently employed. In this section, we have reviewed a variety of quantum educational resources that can support the learners on their journey to enhance their skills and knowledge in the field of quantum technologies.  

\subsection{Academic degree programs}

Understanding of the career opportunities available, as well as the skills, knowledge and education valued by the employers can provide a clear picture of the training and hiring challenges faced by the companies in the quantum industry and the role of higher education in preparing the quantum workforce \cite{PhysRevPhysEducRes.16.020131}. To overcome these challenges, introducing the concepts of quantum in high school will incentivize young people into quantum physics from early on, rather than only discovering the career path late during university physics studies \cite{QBQ}.  Furthermore, developing a roadmap for building a quantum education degree program to incorporate multidisciplinary courses in quantum technologies as standard in degrees such as in computer science, engineering and business or finance, will encourage graduates from other sectors into roles in quantum technologies. Moreover, if the knowledge of quantum concepts, and how they play a role in a field of study are introduced at undergraduate level, then companies will be more inclined to hire bachelor’s or master's graduates rather than PhDs for roles such as quantum engineers or quantum software developers \cite{Qsmart}.

The Quantum Economic Development Consortium (QED-C) conducted a survey to gain a comprehensive understanding of the quantum industry workforce requirements in terms of knowledge, skills, and academic degrees, emphasizing that master's and PhD degrees are prioritized over bachelor's degrees for various jobs in the quantum industry \cite{hughes2021assessing}. With this in mind, it is insightful to analyze the highly-focused quantum degree programs in numerous universities and institutes worldwide. There are currently 162 known universities and institutions worldwide with educational programs and research activities in quantum technologies, according to the Quantum Computing Report \cite{QCR22}. specialized doctoral training programs and institutions, such as the Central Doctoral Training (CDT) programme by University College London (UCL) and the International Graduate School for Quantum Technologies in the UK \cite{UCL, IGSQT}, Quantum Science and Technologies at the European Campus (QUSTEC)\cite{QUSTEC}, the Centre for Quantum Technologies (CQT) in Singapore\cite{CQTS}, and the QuTech Academy in the Netherlands \cite{QUTech}, encourage the development of more programs aimed at preparing the next generation of quantum scientists and researchers. While the doctoral programs offered by the national laboratories and leading universities provide in-depth training and research experiences of specific fields in quantum science and technology, these programs are often expensive and limited at many institutions. Rather than pursuing a specialized PhD degree, many students choose to work directly after graduation. In fact, many industrial professions such as engineers, software developers, and designers require graduates with diverse skills and a basic understanding of quantum physics. It would be beneficial to have more quantum industry focused master's programs to attract graduates to the workforce.

\begin{figure}
\begin{center}

\includegraphics[height=9cm]{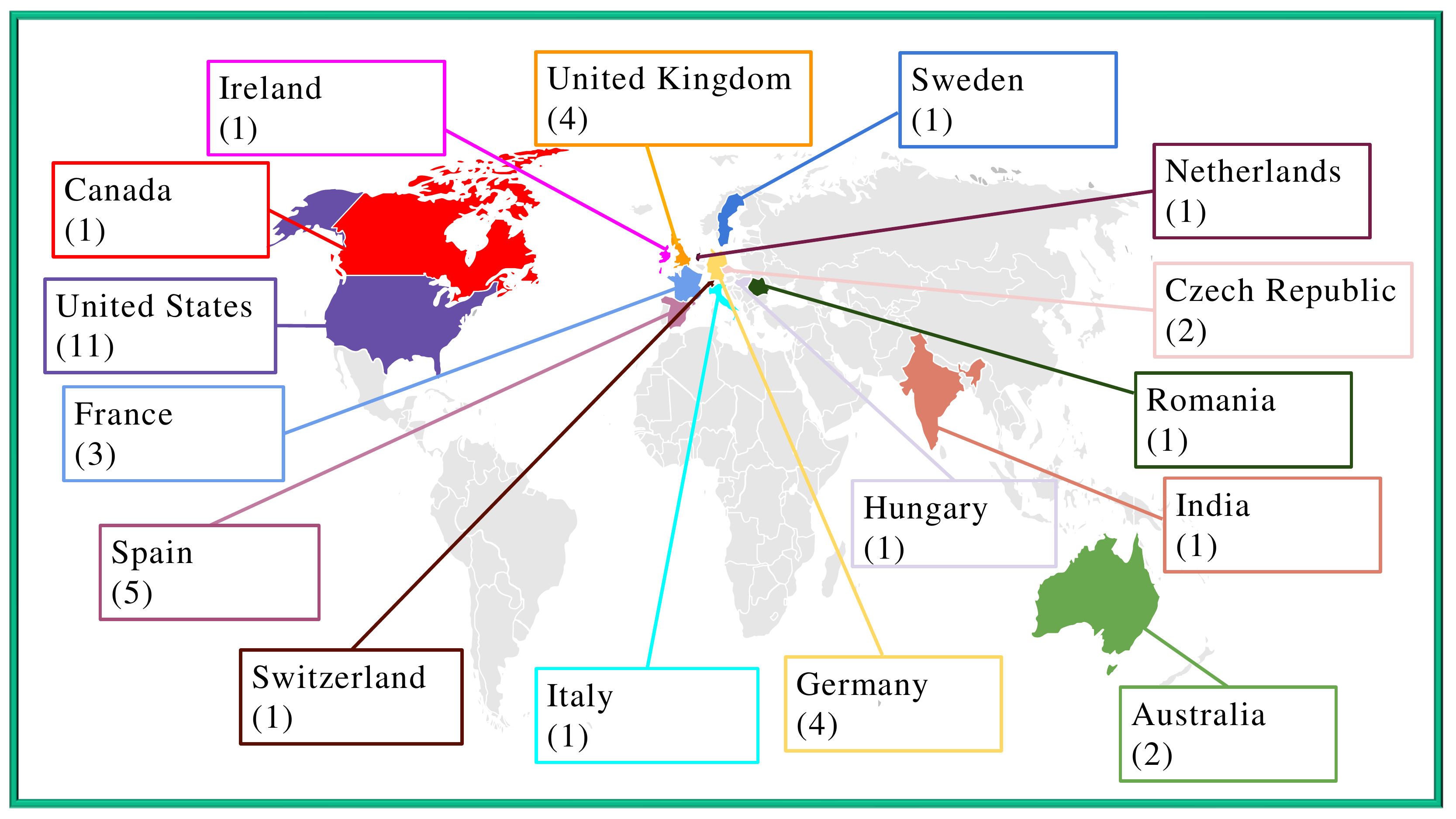}

\end{center}
\caption 
{ \label{fig:4}
Number of master's programs with a focus in quantum technologies across the world. } 
\end{figure}

 QURECA recognized 40 universities worldwide providing master’s degrees in quantum science and technology as shown in Fig.~\ref{fig:4} \cite{Qureca22}. These are mostly concentrated in North America (12) and Europe (21). With 4 Universities in United Kingdom, 2 in Australia and 1 in India, there has been a considerable interest in introducing quantum-related master's programs in a selected group of countries or regions around the world. However, it is highly unlikely that these master's programs would be sufficient enough to meet the needs of the future workforce. Education costs, degree value, and accessibility are some of the limiting factors that can discourage students from enrolling in these programs. Clearly, efforts to create undergraduate and graduate quantum degree programs at a global level must be expanded so that the quantum workforce of the future  is demographically represented and includes all diverse communities. Given the breadth of disciplines — including mechanical engineering, optical engineering, systems engineering, application development and many other areas — that quantum technologies cover, educational institutions should create master's programs that match the criteria of industrial demands. In addition, academic institutions and universities can offer shorter post-graduate certificates or continuing-education programs with the requisite quantum curriculum to meet the growing demand from adults seeking these new skills. For example, the Institute for Quantum Computing (IQC) at the University of Waterloo, Canada has been a pioneer in quantum education, offering interdisciplinary programs to provide knowledge of QIST, including theory and implementations, leading to master's and PhD degree \cite{IQC}. The QTEdu Open Master Pilot project of the Quantum Flagship offers European Transferable Credits (ECTS) that can lead to a master's degree in quantum technology using education programs  across the Europe \cite{OpenMS}. 
 
 Furthermore, the Quantum Flagship’s Quantum Technology Education Coordination and Support Action (QTEdu CSA), gathering data from the European quantum stakeholders, both in industry and in academia, has complied a series of \enquote{Qualification Profiles} under the European Competence Framework to facilitate the planning and design of education and training projects in Quantum Technologies \cite{Qual}, as well as the knowledge skills needed for specific quantum roles in industry. Additionally, many recommendations have been put forward to design the curriculum of these quantum-focused degree programs in order to provide the professional skills that are desirable by the industry at an international level \cite{asfaw2021building, QCCourses, DzurakAus, AminASEE}. Given the administrative and financial constraints of institutions, designing and delivering a bachelor's degree specialized in quantum technologies may be a long road ahead. Institutes can introduce quantum science through current STEM curricula specialized in Physics, Mathematics, Computer Science, and other disciplines \cite{perron}. The primary goals of these programs are to teach existing and recent graduates by delivering a truly multidisciplinary education, as well as to encourage individuals who are currently employed to engage in quantum education. For example, the  training of engineers and programmers as specialists in quantum programming and software can accelerate in order to accelerate the preparation of quantum workforce \cite{peterssen}. While traditional degree programs have always been in demand to fill industry positions, companies are also seeking for applicants who have earned sufficient quantum skills through experience and contributed to the quantum community through alternative educational and training resources.

\subsection{Online courses}

Currently, to be a \enquote{quantum engineer} working on the cutting edge of quantum technological research, one must have a PhD in quantum physics (with very few exceptions) \cite{PhysRevPhysEducRes.16.020131}. This is necessary at the very forefront of the highly technical research, but when the technology comes to fruition, engineers and technicians will be needed too. Some universities are offering more specialized courses geared towards filling this gap in the workforce, but we should also be looking at retraining already qualified engineers, technicians, and business specialists on quantum technologies. Online education providers offer specialized short courses on areas from fundamental quantum science to more specialized  quantum technologies that aim to offer flexible, affordable, job-relevant online learning to individuals and organizations worldwide that can fill the gap needed to build the quantum workforce of today and tomorrow \cite{Qsmart}. 

While edX\textsuperscript{\textregistered}\cite{EdX22} and Coursera\textsuperscript{\textregistered}\cite{Coursera} offer more academic learning focused on a more fundamental approach, QURECA (Quantum Careers and Resources) is the first online training and recruitment platform that provides online courses and resources to fill the gaps in the existing quantum community\cite{Qureca22}. With a mission to create global opportunities in quantum technologies and train the future quantum workforce, QURECA has launched a non-technical course \enquote{Quantum for Everyone}, a specialized education series in Quantum Finance in collaboration with QuantFi \cite{QuantFi}, and its educational platform has been used for global quantum companies such as Zapata Computing \cite{Zapata}. 

The Massachusetts Institute of Technology (MIT)’s Center for Quantum Engineering (CQE) offers online professional development quantum curricula  \enquote{Quantum Computing Fundamentals} and \enquote{Quantum Computing Realities} via MITxPRO to leaders in business, government, and technology who want to learn the core principles, implementations, and implications of quantum technologies \cite{MIT22}. This curricula is designed to bridge the academic gap between quantum science and quantum engineering and thereby advance both. The courses are taught by some of the top quantum computing professors in the world - Isaac Chuang, William Oliver, and Peter Shor.

There are a plethora of quantum computing courses available through online learning platforms such as edX\textsuperscript{\textregistered} \cite{EdX22}, Coursera\textsuperscript{\textregistered}\cite{Coursera}, and Udemy\textsuperscript{TM}\cite{Udm22}. The online education portfolio of Delft University of Technology, Netherlands in edX\textsuperscript{\textregistered} contains a range of certification programs including \enquote{Fundamentals of Quantum Information, Quantum Hardware, Quantum Cryptography, Quantum Software, and Quantum Internet}. Purdue University, USA has released MicroMaster programs in \enquote{Quantum Technology}, specifically in two categories - \enquote{Computing}, and \enquote{Detector and Networking}. In addition, there is an archived \enquote{Quantum Machine Learning} course from Peter Wittek at the University of Toronto, Canada and \enquote{Quantum Mechanics for Scientists and Engineers} course from Stanford, USA available on edX\textsuperscript{\textregistered}. In total, there are 29 courses available on edX\textsuperscript{\textregistered} that provide basic education in quantum technology \cite{EdX22}.

Saint Petersburg State University, Russia offers a \enquote{Quantum Computing: from Basics to the Cutting Edge Specialization} with four courses on Coursera\textsuperscript{\textregistered} for beginner and intermediate level learners, including a course for Russian learners. There are other learning courses such as \enquote{Quantum Mechanics} from University of Colorado Boulder, USA, \enquote{Quantum Optic} from École Polytechnique (France), and \enquote{Exploring Quantum Physics} from the University of Maryland, USA,  accessible through Coursera\textsuperscript{\textregistered}\cite{Coursera}. \enquote{Understanding Quantum Computers} developed by Keio University, Japan, is a short course available on FutureLearn for anyone interested in quantum computing at the \enquote{popular science} level\cite{UQCf}. With the ability to run code on quantum computers, qBraid is offering a learning platform for programming as well as introductory courses in quantum computing targeted at different education levels \cite{qbraid22}. \enquote{Quantum Computing For The Very Curious} by Andy Matuschak and Michael Nielsen from Quantum Country came up with a mnemonic approach for learners with a focus on describing all the basic principles of quantum computing and quantum mechanics, and the quantum search algorithm and quantum teleportation in more detail \cite{QCftVC19}.

With an aim to empower a quantum community of lifelong learners, address challenges and opportunities, and sustain the quantum initiative in the long-term, the Institute of Electrical and Electronics Engineers (IEEE) Quantum's new flagship \enquote{Quantum Computing Education - Workforce Development Program} is offering courses in various disciplines in quantum technology presented by industry leaders \cite{IEEE21}. Q-CTRL, an Australia-based quantum computing startup, introduced Black Opal, an interactive platform for beginners designed to provide training to software engineers, security analysts, and data scientists in the fundamentals of quantum computing \cite{Q-CTRL}. Microsoft\textsuperscript{\textregistered} collaborated with Alphabet X and California Institute of Technology's Institute for Quantum Information and Matter (IQIM) to release a quizzes-based course in quantum computing on Brilliant \cite{Brilliant22}. Apart from these, there are other online learning resources available such as  quantum computing courses on Udemy\textsuperscript{TM} \cite{Udm22}, multimedia learning resources on QPlayLearn \cite{QPLearn}, and articles on GeeksforGeeks\textsuperscript{TM}, a computer science portal \cite{G4G}.

\subsection{Conferences, workshops, and hackathons}

On the journey to build a stable, inclusive, and diverse quantum workforce, substantial efforts are needed to reduce the barriers to lifelong learning and make quantum education open and accessible to everyone in the world. This can be done through inclusion of educational and professional development activities such as workshops, summer schools, and community-building events. 

IBM Quantum initiated the most prominent Qiskit Global Summer school with a focus to empower the next generation of quantum researchers and developers, and impart knowledge in quantum computing, quantum machine learning, and quantum error correction globally \cite{Qiskit21}. The curriculum covered in these summer schools have been transformed into online courses. In addition, an open-source Qiskit textbook has been used in universities around the world as quantum algorithms or quantum computation course supplement to teach quantum computation using Qiskit Software Development Kit (SDK) for working with quantum computers. The Institute for Quantum Computing (IQC) has been conducting summer schools and seminars for students ranging from high school to graduate level through world-class outreach programs. Aimed at providing a comprehensive introduction in quantum technologies to undergraduate and master's students we can find initiatives such as the Quantum Computing Summer School by Los Alamos National Laboratory (LANL) \cite{LANL} , the European Spring School in Quantum Science and Technology \cite{ESQST}, the UCLQ Quantum Tech School \cite{UCL}, the Munich Center for Quantum Science and Technology (MCQST) summer student program \cite{MunichSch}, and the Quantum Computing Summer School Camp by SQA \cite{SQA}. A \enquote{Quantum Summer School} by QWorld \cite{qworld22} is another example of initiatives addressed at students, researchers, or any professional interested in a career development towards quantum computing.

Conferences can become excellent opportunities to network with other professionals with similar interests and to learn new trends, ideas and solutions. Conferences in quantum technologies allow to:

\begin{itemize}

    \item bring together quantum professionals, researchers, educators, entrepreneurs, champions and enthusiasts to discuss the potential of quantum technologies,
    
    \item address recent advances in infrastructure development, education and training, and groundbreaking efforts in scientific research and development,
    
    \item exchange and share the experiences, challenges, research results, innovations, applications, pathways and enthusiasm on all aspects of quantum computing and engineering,
    
    \item promote communications between different stakeholders, industry leaders, researchers, engineers, scientists and general public, 
    
    \item and offer networking opportunities and industry engagements to foster the global quantum ecosystem.
    
\end{itemize}

Some of the noteworthy conferences being held globally (or virtually) to bring together the quantum community include the European Quantum Technologies Conference (EQTC) by the Quantum Flagship \cite{EQTC21}, the IEEE Quantum Week - International Conference on Quantum Computing and Engineering (QCE) \cite{IEEEQW}, the conference on Quantum Information Processing (QIP) \cite{QIPconf22}, the Quantum 2.0 Conference and Exhibition organized by Optica (formerly OSA) \cite{OpticaConf22}, the American Physical Society (APS) Meeting \cite{APS22}, SPIE Quantum West \cite{Spie22}, and QCrypt \cite{Crypt22}. To accelerate the commercialization of quantum applications in the industry, there are now several business-focused events such as Quantum Business Europe (QBE) \cite{QBE}, Practical Quantum Computing (Q2B) \cite{Q2B}, Inside Quantum Technology \cite{iqt22}, Commercialising Quantum \cite{EcoQu22}, and Quantum Tech \cite{alpha22}, bringing together industry leaders, start-ups, marketing strategists, early adopters, and government officials, bridging the gap between research and industry.  

QURECA aims to educate the quantum community by providing opportunities to engage and network with industry leaders through these conferences and events, and by bringing together diverse stakeholders to advance the global quantum ecosystem. With this in mind, QURECA launched Quantum Latino \cite{Latino22} and Quantum Eastern Europe \cite{qee} to promote quantum communities in Latin America and Eastern Europe respectively, with the objective of creating more community-based events in the near future. The Quantum Computing Lab initiative by CINECA has been organizing the High Performance Computing and Quantum Computing (HPCQC) workshop since 2018 acknowledging this new technological revolution \cite{QuantCompLab}. In addition, Quantum Computing Theory in Practice (QCTIP) is another workshop held in the UK to foster discussions between theorists and practitioners of quantum computing \cite{QCTIP22}. 

Quantum computing industry leaders and start-ups, in collaboration with universities and research institutes are taking the initiative to conduct quantum computing hackathons with an opportunity for learners to solve real world-problems by programming and simulating real quantum machines. Recent quantum computing hackathon initiatives include QHack by Xanadu\cite{Qhack22}, the Airbus Quantum Computing Hackathon \cite{Air22}, the Big Quantum Hackathon by QuantX \cite{QuantX22}, the Quantum Futures Hackathon by Quantum Open Source Foundation (QOSF) \cite{Qosf}, MIT iQuHACK \cite{iquhack22}, Quantum Coalition Hack \cite{QuantColl22}, and QPARC Challenge \cite{QPARC22}. These hackathons allow participants from different backgrounds to learn about quantum technologies, identify quantum applications in various industries, develop their own portfolio projects using quantum computing cloud services, and network with engineers, researchers, academic and industry experts.

\subsection{Games}

A promising new frontier in educating the quantum principles, often considered confusing or counter-intuitive, is the use of game-based learning. Incorporating the visual clues to understand the core principles of quantum science make games a powerful tool in engaging learners to build their analytical thinking, multitasking, strategizing, problem-solving, and team-building skills, which consequently help in addressing the challenge of quantum literacy \cite{QuLiteracy}. 

IBM Hello Quantum designed by IBM Research in collaboration with Professor James Wootton at the University of Basel (Switzerland) is a puzzle game designed for non-expert in building intuition and knowledge on how to apply quantum computing principles in coding quantum programs by visualizing just 2 qubits \cite{IBMhel22}. It provides a gateway to explore and experiment on IBM Q Experience. Quantum Chess is like an ordinary chess, but with a quantum twist to introduce an element of unpredictability \cite{QuantChess22}. In Quantum Chess, rules are modified to incorporate quantum principles like superposition and entanglement. Quantum Flytrap has created a virtual lab for playing games with photons to make undergraduate physics experiments more enjoyable and interactive. With a drag and drop interface, it helps learners to visualize quantum phenomena, recreate existing, and prototype new experiments \cite{Quantflytrap22}. With an aim to inform and educate people in quantum technologies, science at home launched Quantum Moves 2 in helping researchers to solve and optimize the manipulation of qubits and make their contribution in cutting-edge quantum research \cite{QuantGames22}. The Institute of Quantum Computing(IQC) at University of Waterloo, Canada, unveiled Quantum Cats, an Angry Birds-inspired game that highlights quantum behavior and the differences between quantum physics and classical physics \cite{Quantumcats22}. Quantum Odyssey is the first video game to help learners  immerse themselves in the world of quantum computing \cite{QOdyssey}. The Quantum Mechanics Visualization Project (QuVis) developed by the University of St. Andrews, Scotland, UK designed a research-based interactive simulations for learning and teaching quantum mechanics concepts \cite{QuVis}. 

Even beyond a virtual platform, Spin-Q Gemini offers a quantum computing desktop platform, launched to facilitate hands-on training and real-device experience in understanding quantum computing to K-12 and college students \cite{SpinQ}. Understanding quantum mechanics may seems intimidating at first, but employing games as an educational tool ensures that quantum enthusiasts are engaged in fun and engaging learning while also broadening their skills. 

\subsection{Community-building and other resources}

In an effort to accelerate quantum workforce development, many organizations have institutionalized a variety of education and training practices by incorporating content beyond traditional course learning and education tracks. To build inclusive and diverse quantum communities, QURECA \cite{Qureca22}, QWorld\cite{qworld22}, and OneQuantum \cite{onequantum22}  have established global quantum groups for creating awareness about quantum technologies by enhancing student engagement through workshops, mentoring, career and networking events, and outreach activities. These organizations are collaborating with academics and industry experts to identify, nurture, progress, and retain the next generation quantum workforce. By facilitating quantum education through e-learning platform, SheQuantum is empowering women to contribute towards talented quantum workforce \cite{SheQuantum}. The Unconventional Computing Lab \cite{Unconventional}, Brazil Quantum\cite{BrazilQ}, and aQuantum \cite{aquantum} are educating and promoting quantum technologies in various communities by reducing the language barriers. Qubit by Qubit initiative by The Coding School is aimed at educating high school and undergraduate students through hands-on and innovative programs, workshops and courses in the field of quantum computing \cite{QBQ}. Quantum Curious, a platform featuring a range of learning resources for anyone interested in quantum computing, is a great way to attract a wider audience without a strong STEM background \cite{QuantCur}. With a mission to support the development and standardization of open tools for quantum computing, the Quantum Open Source Foundation (QOSF) \cite{Qosf}, The Quantum Insider's integrated platform Entangle \cite{QEntangle}, Quantum Computing Stack Exchange \cite{QCSE}, and Quantum Computing powered by Strangeworks \cite{Strangeworks22} are examples of platforms providing a means of learning quantum computing, coding and sharing open-source quantum software projects with the global quantum tech community. Furthermore, addressing the current landscape of quantum technologies, exchanging knowledge and innovation in research, leveraging existing initiatives and collaboration, and supporting the quantum community through curated quantum education web resources for students, teachers, and the general public form the basis of major initiatives that recognize  quantum technology's potential to revolutionize research \cite{AAPT, CERN22, IEEE21, IOP}.  Raising awareness and insight into the future impact of quantum technology through a series of lectures, volunteering, presentations or symposiums as well as the prospect of career opportunities could motivate people to venture into underserved fields of quantum technology \cite{QPARC,QuSTEAM, FinQ, QuSoft}.

\begin{figure}
\begin{center}

\includegraphics[height=9.0cm]{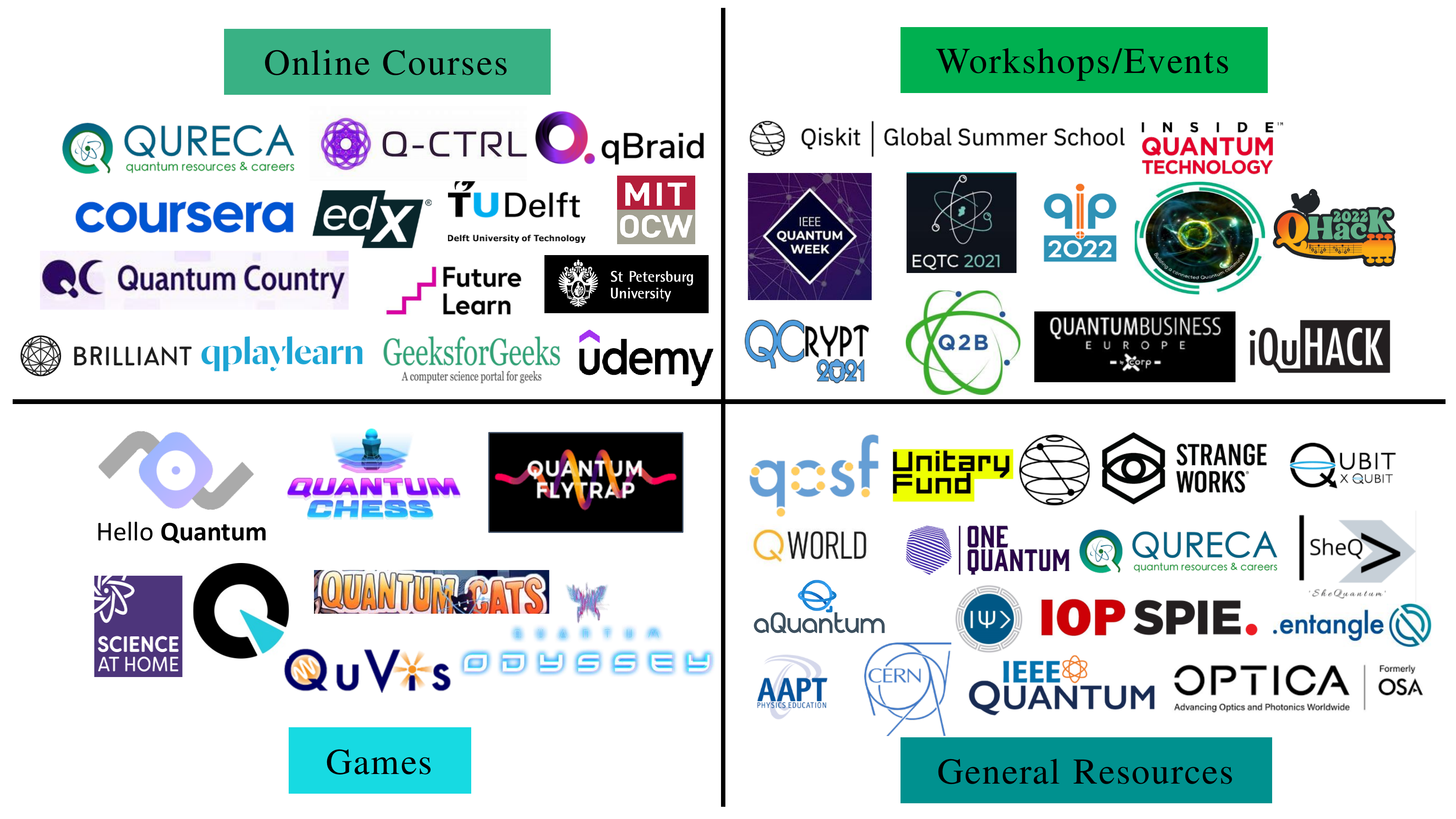}

\end{center}
\caption 
{ \label{fig:5}
Overview of global educational and training resources available for the development of the current and future workforce. } 
\end{figure} 

We have summarized all kinds of educational resources including learning platforms offering online courses, conferences, workshops and hackathons, games, and other community-building platforms in the field of quantum technologies in Fig.~\ref{fig:5}. Please keep in mind that this is a comprehensive list of accessible education and training resources. As the second quantum revolution unfolds, numerous new businesses, organizations, academic and private institutions have expressed an increased interest in deepening their involvement in this rapidly expanding field.

\section{Industry workforce strategy}\label{section6}
\begin{figure}
\begin{center}

\includegraphics[height=9.0cm]{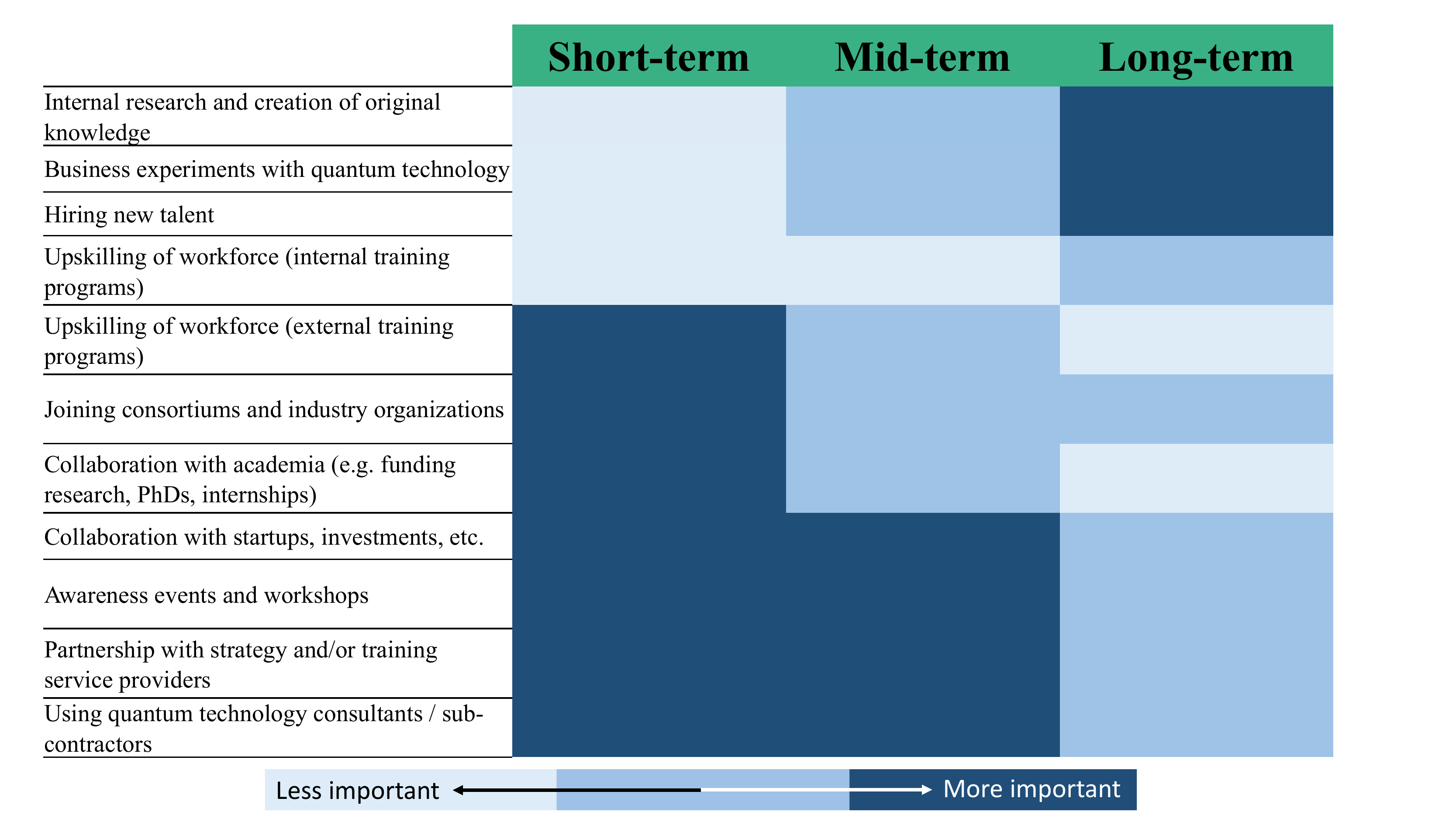}

\end{center}
\caption 
{ \label{fig:6}
The components of a strategy to create a quantum-ready workforce (short, medium, and long-term) in the current industry landscape.}
\end{figure}

The landscape analysis of current employment needs and available training programs (offered through traditional education degrees or other education initiatives) is key to identify the underlying drivers of the skills mismatch, to design strategies for training and upskilling the quantum workforce. In the numerous quantum events and discussions focused on the business aspect, rather than the academic's, of quantum technologies, one of the most common requests regarding education and workforce in the current landscape comes from individuals and industry partners, i.e. end users. They want to know how to get started with quantum technologies, what are the resources available, and how they could develop their own strategy. Depending on their quantum awareness level, whether the individuals are technical professionals or have expertise in QIST\cite{qistUS}, or are completely new to the field, a plan with short, medium, and long-term activities will support the creation of their own workforce. In Fig.~{\ref{fig:6}} we provide an overview of the different activities that can be undertaken by the current industry stakeholders (end users) in the short, medium and long-term, by prioritizing certain actions to address the workforce needs.

\section{Conclusion}\label{section7}

The rapid expansion of quantum technologies needs the development of a well-rounded quantum workforce strategy that encompasses the quantum education at all levels, and embraces the skills and training to produce quantum-ready talent across all academic, gender, racial, and ethnic backgrounds. This paper has presented an assessment of the education and training resources available worldwide, which are critical for the development of the existing and future quantum workforce, as well as the educational and career paths possible in the current ecosystem.

 The quantum industry is geared up in transforming the research ideas to real-word applications and products and this transformation will not be accomplished by recruiting just PhD-level quantum physicists. Instead, enterprises should consider hiring additional professionals such as engineers, programmers, product designers, strategists, and others. We observed that the current workforce prioritized PhD degree candidates as compared to bachelor's and master's, and that higher education demands were more prominent in Europe and rest of the world than in the US or Canada. In conclusion, academic degrees only cannot provide industry desirable job-ready skills. Industry experience is sometimes much more valuable than general academic degrees, and by upskilling the current workforce we can tackle the identified quantum skills bottleneck.
 
 To create and strengthen the pipeline between academia and industry, there are several paths that integrate disparate educational resources into rich, diverse, cohesive and informative learning experiences. Incorporating quantum skills and training through a designated sequence of activities, often from different sources and dedicated programs, can educate the quantum talent of tomorrow. We outlined various learning resources such as online courses, conferences, workshops, hackathons, games, and community-building forums. Each of these resources has a varying level of accessibility addressing the different needs of the current and future quantum workforce. 
 
 Furthermore, we described the activities that can be undertaken by industry to develop their quantum strategy and workforce. This is directly linked to the short term solution to create quantum talent, which is to develop the skills of the current professionals by initiatives focused on retraining the existing workforce. Master's and PhD programs will address the workforce at a longer term, and several of these initiatives have been mentioned in the paper. For a much longer term, the education of current high school and primary school students is key. Although some of the resources mentioned in the manuscript address this audience, the detailed overview of education initiatives at this level falls beyond the scope of this paper. Comprehending the available resources and outlining the potential strategies and paths are key to bridge the quantum knowledge gap in the workforce.

%%%%% Acknowledgments %%%%%
\textit{Acknowledgements}

The authors gratefully acknowledge the discussions with our industry partners, and the quantum community across academia, government, and national labs. We also thanks to Kirsty Morrison and Andreas Landsmann for their immense contribution, cooperation, and help during the preparation and review of this paper. 

%%%%% References %%%%%

\bibliography{report}   % bibliography data in report.bib
\bibliographystyle{spiejour}   % makes bibtex use spiejour.bst

%%%%% Biographies of authors %%%%%

\vspace{2ex}\noindent\textbf{Maninder Kaur} received her PhD degree in theoretical physics from University of Delhi, India in 2018. She worked as a theoretical plasma physicist in LUIS experimental team in ELI Beamlines, Czech Republic. Her research focused on numerical modeling and computational simulations in Laser Plasma Accelerators and nonlinear optics. Her passion in exploring the next-generation technologies in research and applications using machine learning and quantum computing inspired her to work at QURECA.

\vspace{1ex}

\vspace{2ex}\noindent\textbf{Araceli Venegas-Gomez}, took a career change from
being an aerospace engineer at Airbus in Germany to become a quantum physics researcher at the University of Strathclyde in Glasgow, Scotland. In 2019, she was awarded a fellowship by the Optical Society for her contributions to the
public and business awareness of quantum technologies. She founded QURECA to build a bridge between research and industry, supporting business
and institutions to join the quantum revolution.

\vspace{1ex}

\listoffigures

\end{spacing}
\end{document}